\documentclass[12pt,english,reprint]{elsarticle}
\usepackage[LGR,T1]{fontenc}
\usepackage[latin9]{inputenc}
\usepackage{graphicx}

\newcommand*\LyXThinSpace{\,\hspace{0pt}}

\usepackage{babel}
\usepackage{hyperref}

\newcommand{\be}{\begin{equation}}
\newcommand{\ee}{\end{equation}}
\newcommand{\ba}{\begin{eqnarray}}
\newcommand{\ea}{\end{eqnarray}}

\makeatother

\usepackage{babel}
\begin{document}

\title{High-quality GeV-scale electron bunches with the Resonant Multi-Pulse
Ionization Injection}

\author{P. Tomassini$^{1}$, S. De Nicola$^{2,6}$, L. Labate$^{1,3}$,
P. Londrillo$^{4}$, R. Fedele$^{5,6}$, D. Terzani$^{1,5,6}$, F. Nguyen$^{7}$ , G. Vantaggiato$^{1}$, and L. A. Gizzi$^{1,3}$ }

\address{$^{1}$Intense Laser Irradiation Laboratory, INO-CNR, Pisa (Italy)}


\address{$^{2}$SPIN-CNR, Sect. of Napoli, (Italy)}

\address{$^{3}$INFN, Sect. of Pisa, (Italy)}

\address{$^{4}$INAF, Bologna (Italy)}

\address{$^{5}$Dip. Fisica Universita' di Napoli Federico II (Italy)}

\address{$^{6}$INFN, Sect. of Napoli (Italy)}

\address{$^{7}$ENEA, Nuclear Fusion and Safety Technologies Department, Frascati (Italy)}

\begin{abstract}
Recently a new injection scheme for Laser Wake Field Acceleration,
employing a single 100-TW-class laser system, has been proposed. In
the Resonant Multi-Pulse Ionization injection (ReMPI) a resonant train
of pulses drives a large amplitude plasma wave that traps electrons
extracted from the plasma by further ionization of a high-Z dopant
(Argon in the present paper). While the pulses of the driver train
have intensity below the threshold for the dopant's ionization, the
properly delayed and frequency doubled (or more) ionization pulse
possesses an electric field large enough to extract electrons, though
its normalized amplitude is well below unity.   In this paper
we will report on numerical simulations results aimed at the generation
of GeV-scale bunches with normalized emittance and {\it rms} energy
below $80\, nm \times rad $ and $0.5\, \%$, respectively. Analytical consideration of the FEL performance for a $1.3\, GeV$ bunch will be also reported.  
\end{abstract}
\maketitle

\section{Introduction}

The Resonant Multi-Pulse Ionization injection (ReMPI) scheme is derived from
the so-called ``two-color ionization injection''. In the two-color
ionization injection \cite{TWOCOLOUR,TWOCOLOUR2} two laser systems
are needed. The main pulse that drives the plasma wave has a long
wavelength, five or ten micrometers, and a large normalized amplitude
$a_{0}=eA/mc^{2}=8.5\cdot10^{-10}\sqrt{I\lambda^{2}}>1$, being $I$
and $\lambda$ pulse intensity in $W/cm^{2}$ and wavelength in $\mu m$.
The second pulse (the ``ionization pulse'') is a
frequency doubled Ti:Sa pulse with wavelength $400\, nm$. While the main pulse cannot further
ionize the electrons in the external shells of the large Z
dopant due to its large wavelength, the electric field of the ionization
pulse is large enough to generate newborn electrons that will be trapped
in the bucket. This opens the possibility of using gas species with
relatively low ionization potentials, thus enabling separation of
wake excitation from particle extraction and trapping. Two color ionization
injection is therefore a flexible and efficient scheme for high-quality
electron bunch production. The main drawbacks of the two color ionization
injection are the current lack of availability of short (T<100 fs)
100 TW-class laser systems operating at large ($\approx 10\,\mu m$) wavelength
and lasers synchronization jitter issues. These limitation make the
two-color scheme currently unpractical for application to LWFA-based
devices requiring high quality beams. 

The Resonant Multi-Pulse Ionization injection \cite{RMPII} has the
possibility to be operating with present-day {\it single} Ti:Sa laser
systems. Simulations show that such a scheme is capable of generating
ultra-low emittance GeV-scale bunches with easily tunable length and
final energy.

\section{The Resonant Multi-Pulse Ionization Injection}

In the Resonant Multi-Pulse ionization injection scheme (see Fig.
1) only one short-wavelength laser system (e.g a Ti:Sa) is needed.
The long wavelength driving pulse of the two-color scheme is replaced
by a short wavelength, resonant multi-pulse laser driver. Such a driver
can be obtained via temporal shaping techniques from the \textit{single},
linearly polarized, standard CPA laser pulse, while the minor fraction of the Ti:Sa CPA pulse is frequency doubled and used as an ionizing pulse.  

\begin{figure}
\includegraphics[scale=0.3]{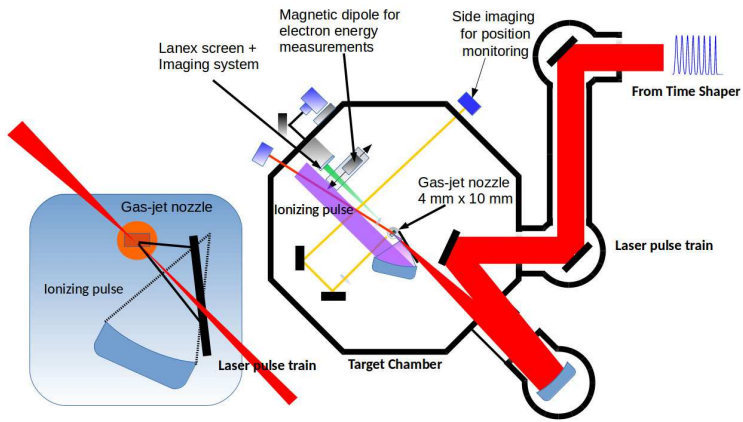} \caption{Multi-Pulse ionization injection scheme. A small fraction of a single
Ti:Sa laser pulse is frequency doubled and, after focusing with a
low F/\# paraboloid, will constitute the ionizing pulse. The main portion
of the pulse is temporally shaped as a train of resonant pulses that
will drive a large amplitude plasma wave. Inset: the ionizing pulse
focusing is achieved by using a mirror with a hole for the driving
pulse passage. }
\end{figure}

 Due to the resonant enhancement of the ponderomotive
force, a properly tuned train of pulses is capable of driving amplitude
waves larger than a single pulse with the same energy \cite{MULTIPULSE,MULTIPULSE2}.
Noticeably, since the peak intensity of the driver is reduced by a
factor equal to the number of train pulses, it is also possible to
match the conditions of \textit{both} particle trapping and unsaturated
ionization (i.e. with low ionization percentage) of the active atoms level. Recently \cite{MP-EXP} new
experimental results on the generation of such a time shaped pulses
demonstrate that a multi pulse scheme is obtainable with present day
technology and that plasma waves can be excited with this scheme \cite{MP-PRL}.
Using Argon ($Ar^{8+}\rightarrow Ar^{9+}$ with ionization potential
$U_I=422.5\, eV$) as a dopant gives us the possibility to obtain
bunches with tens of $nm\times rad$ of normalized emittance. Multi-pulse ionization
injection with Argon requires trains with more than four pulses since
ionization level is saturated  with amplitude above $a_{0}=0.8$ at
$\lambda=0.8\,\mu m$ (see Fig. 3 in \cite{RMPII}).

\section{1.3 GeV beam simulation }

We report on a long acceleration length (of about 4 cm) simulation 
performed in a 2D cylindrical geometry with QFluid \cite{QFLUID}(see also the Appendix in \cite{RMPII}). The Ti:Sa laser system generates pulses that will pass through a beam
splitter. The major portion of each pulse is time shaped as a train
of resonant eight sub-pulses having FWHM duration of $T=30\, fs$ each, with
peak power of 200/8 TW. The driving train is subsequently focused down to a
spot of $w_0=45 \mu m$ waist onto a capillary filled with Argon, obtaining a sequence of pulses with peak intensity and normalized amplitude of $I = 7.9\times 10^{17}W/cm^2$ and $a_0=0.6$, respectively.
The frequency doubled pulse from the minor portion of the Ti:Sa pulse
delivers $13\, mJ$ and is focused with a minimum waist of $w_{0, ion}=3.6 \mu m$.
On-axis plasma background density is set to $ n_{axis}=5\times 10^{17} cm^{-3}$
and is obtained by assuming full ionization of Argon up to level eight
(ionization potentials of $Ar^{n+}$ are below $144\, eV$ for $n \le 8$
so Argon ionization up to $Ar^{8+}$ is achieved within the first
cycles of the pulse).

To obtain a so long acceleration length pulse guiding technique is
necessary since low-density plasmas don't allow for pulse self-guiding
at those pulse powers. The driver pulses are focused close to the
entrance of the capillary (or gas-cell) and enter into the guide with
a matched radius $w_m=w_0$ and radial density profile 

\begin{equation}
n_{e}(r)=n_{axis}\left[1+\eta\frac{1.1\cdot10^{20}}{n_{axis}w_{0}^{2}}\left(\frac{r}{w_{0}}\right)^{2}\right]\,.
\end{equation}
 The $\eta$ factor accounts for weakly nonlinear corrections and
in the case of short pulses ($T<<2\pi/\omega_p$  can be evaluated
as \cite{GUIDING} $$\eta\cong1-\frac{1}{16}(a_{0}\omega_{p}T)^{2}\cdot\left(1+4.6\cdot10^{-21}n_{e}w{}_{0}^{2}\right)\, ,$$
which is very close to unity in our simulations. 

Simulation has been performed onto a moving cylinder of radius $160 \, \mu m$, length $430\,  \mu m$ and a resolution in both radial and longitudinal
directions of $150\, nm$. Due to the tight focusing of the ionization pulse that diffracts in a scale $Z_{r,ion}=\pi\times w_{0,ion}^2/\lambda_{ion}\simeq 100\, \mu m$, the bunch population growths and saturates (bunch charging phase) in  about $150\, \mu m$
(see Fig. 2, green dots representing the longitudinal phase-space
of the bunch in the charging phase) and the extracted bunch is trapped
after $\approx 600\, \mu m$ of propagation of the ionizing pulse (see red dots in Fig.
2) in a phase of the bucket intermediate between the weak-trapping
and the strong-trapping conditions (see Eqq. 2 and 3 in Ref. \cite{RMPII}). 

\begin{figure}[h]
\includegraphics[scale=0.55]{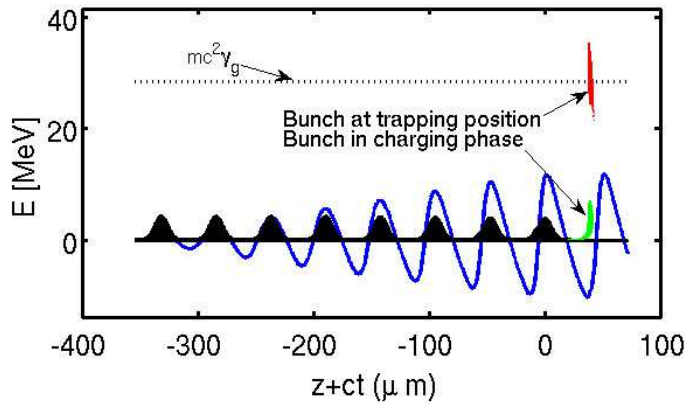}\caption{Line-out of the longitudinal electric field (blue line, a.u.) and
pulse amplitude of the driver train at the early stage of bunch trapping.
Green dots show the longitudinal phase space of the bunch after $100\, \mu m$
of propagation. The horizontal dotted line shows the energy at the
trapping point ($\gamma_g$ is the Lorentz factor of the pulse train)
and the red dots represent the longitudinal phase-space of the bunch
at the trapping point (i.e. $<\gamma>=\gamma_g$)}
\end{figure}

The driver pulses evolution through the $3.7\, cm$ of plasma shows a twofold behavior.
Though peak intensity is remarkably stable (see the black line in
Fig. 3), and no visible self-steepening occurs (we are well below the threshold for the onset of self-steepening since $a_0 \times (c\times T)\times k_p\approx 0.8$ and according to \cite{SelfSteepening} the growth of self-steepening occurs if $a_0 \times(c\times T)\times k_p>\left(32\times log(2)/(\pi-1)\right)^{1/2}\approx 3.2 $)   sub-pulses of the rear part of the train propagate in the
wake generated by all the preceding pulses, thus being partially exposed
to the defocusing effect of the wake. As a final effect, a radial
breathing of the rear pulses occurs with possible off-axis maxima
of the local intensity, as it is apparent in Fig. 4 (bottom).

\begin{figure}[h]
\includegraphics[scale=0.55]{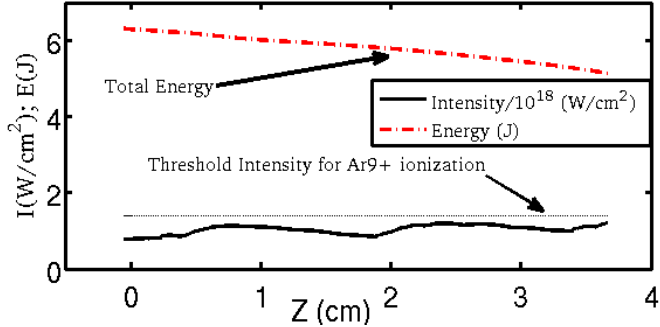}\caption{Evolution of the total energy (red line) an peak intensity (black
line). The horizontal dotted line represents the intensity threshold
for further ionization of the 9-th level of Argon. }
\end{figure}

The final electron bunch of charge $4.3\, pC$ has energy 1.3 GeV,
energy spread 0.49\% {\it rms} and normalized emittance of $0.08\, mm\cdot mrad$
and $0.04\, mm\cdot mrad$ in x (laser polarization) and y directions,
respectively. After 3.7 cm of propagation the electron bunch is still
far from dephasing (see Fig. 4 top) and almost 70\% of laser energy
is still available for further energy boost. However, while normalized
emittance looks stable in the last $3\, cm$ (see Fig. 5) due to the matched-beam
configuration, the relative energy spread finds its minimum at $3.7\, cm$
and rapidly increases with further acceleration up to percent level.
For high-quality oriented application, therefore, such a earlier truncation
of particle acceleration limits the overall energy conversion efficiency
of the scheme (at the present working point). We finally stress the
remarkably low value of $0.2\,\%$ for the slice energy spread (with slice
thickness of $0.05\, \mu m$).  

\begin{figure}
\includegraphics[scale=0.55]{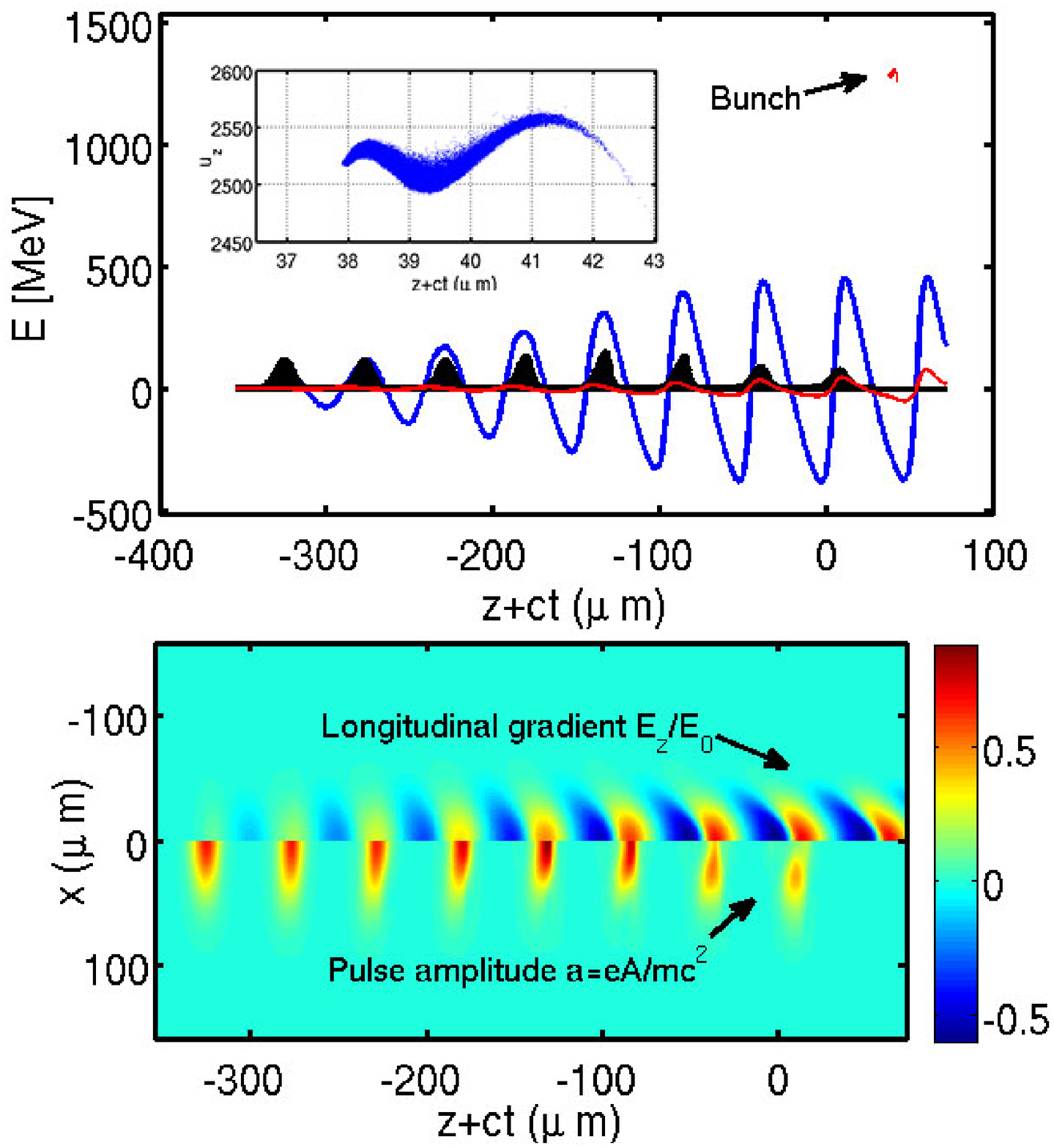}\caption{Top: longitudinal phase space of the electron bunch after $3.7\, cm$ of
propagation (red dots). The blue line shows the electric field on
axis (a.u.), while the red line represents the transverse focusing
force at a radius close to the beam radius (a.u.). Bottom 2D maps
of the longitudinal normalized electric field $E_z/E_0$ and of the
normalized laser amplitude.}
\end{figure}

\begin{figure}
\includegraphics[scale=0.5]{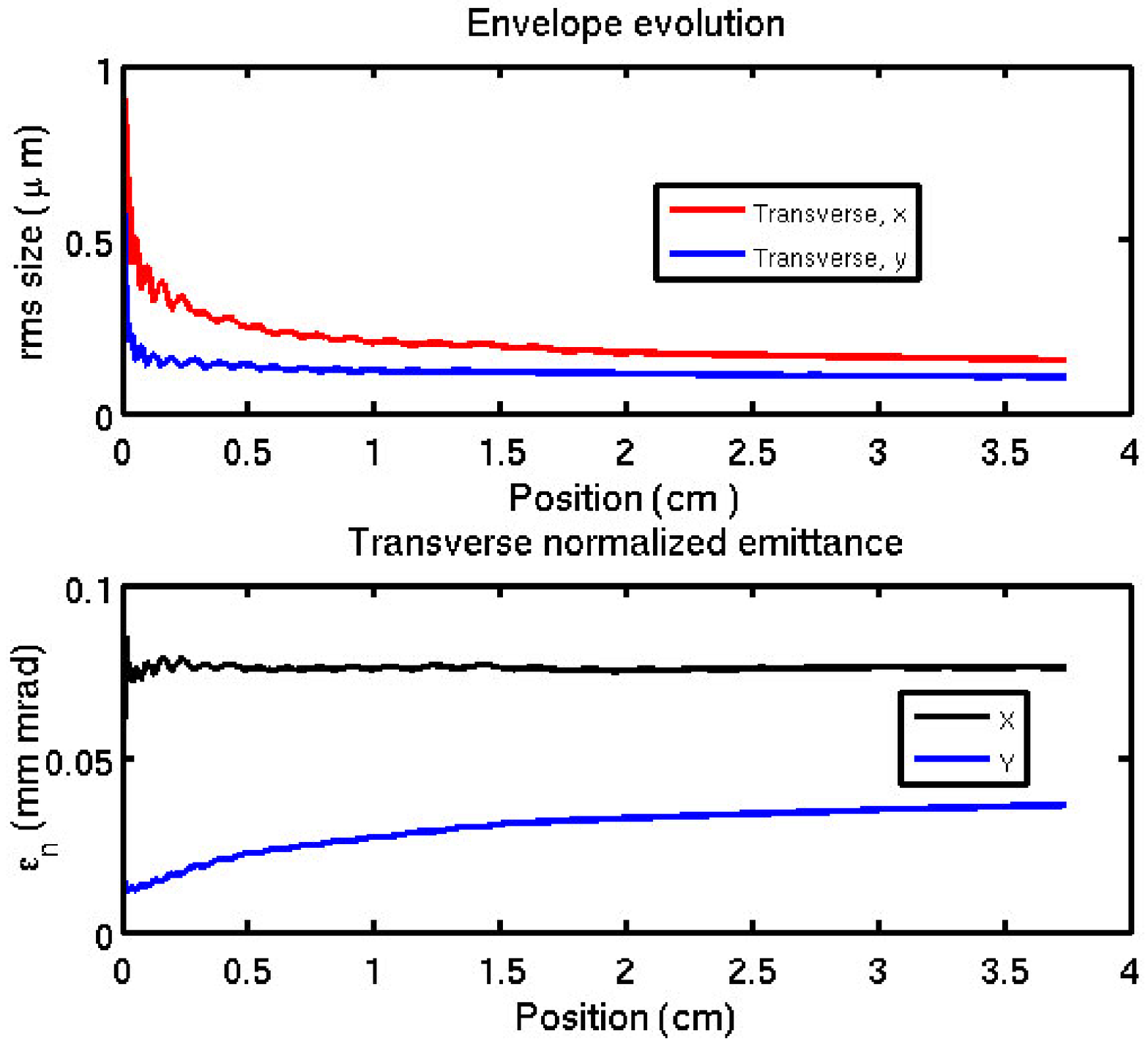}

\caption{Transverse rms size (top) and normalized emittance (bottom) in $x$ (pulse
polarization) and $y$ directions. }

\end{figure}

The simulation ends when the pulses and the bunch are close to the plasma exit. Due to the use of the quasistatic approximation, QFluid cannot face with rapidly varying longitudinal plasma densities so we will be forced to use a different code to face with the plasma exit stage. 

The ReMPI scheme uses a single laser system (a Ti:Sa in the present
paper) so the driving train and the ionization pulse have no relative
timing jitter. This opens the possibility of fine tuning the ionization-to-driver
delay according to the requested bunch energy or length.
 The fine-tuning of the bunch duration is easily obtainable just by selecting the appropriate ionization-to-driver
delay $t_d$. Numerical simulations (supported by theory in 1D) show
that the minimum bunch length is obtained when the ionization pulse
is placed at the position of maximum potential (zero longitudinal
electric field) and the trapped bunch is placed at the position of
maximum accelerating gradient (i.e. exactly at the strong-trapping
point). Starting from that configuration and further delaying the
ionization pulse, the final bunch length increases. The fulfillment
of the weak-trapping condition for the whole set of bunch electrons
makes an upper limit of the bunch duration. Both the minimum and maximum
obtainable values depend on the working point. In the current setup
bunch lengths that can be obtained by simply delaying the ionization
pulse are in the range $360\, as < t_{rms} < 2.2 fs$. Optimization
of the bunch length/energy tuning strategy is ongoing \cite{InPrep}.

\section{Estimates of FEL performance}
In this section we will report on analytical results obtained with the $1.3\, GeV$ bunch, having supposed an emittance preserving beamline to transport the bunch from the accelerator stage to the undulator. We stress, therefore, that the following results do not constitute the final stage of a start-to-end simulation but just the expected outcomes of FEL radiation in the case of a quality-preserving optics. 

For an electron beam of energy $E_{beam}=\gamma m_e c^2$, the resonance condition for the wavelength of the emitted radiation, in a planar undulator, is
\be
\label{scaling:1}
\lambda = \frac{\lambda_u}{2\gamma^2}\left(1+\frac{K^2}{2}\right)
\ee
where $\lambda_u$ is the period of longitudinal variation of the on-axis magnetic filed for a planar undulator and K is the undulator parameter defined as:
\be
\label{scaling:2}
K=\frac{e B\lambda_u}{2\pi m c}
\ee
$B$ being the peak value of the on-axis magnetic field and $e,$ $m_e$ and $c$ respectively the electron charge, the electron mass and the speed of light. 

\begin{table}[htp]
\caption{Performance estimates of a Free Electron Laser driven by the electron beams discussed in the text.\label{tab:fel}}
\begin{center}
\renewcommand{\arraystretch}{1.4}
\begin{tabular}{|l|c|c|c|}
\hline\hline
\multicolumn{4}{|c|}{\em Bunch parameters}\\
\hline
beam energy [GeV] & \multicolumn{3}{c|}{$1.3$}\\
\hline
long. beam size (rms) $\sigma_L$ [$\mu$m]  &  \multicolumn{3}{c|} {$0.655$}\\
\hline
current intensity [A]  &  \multicolumn{3}{c|} {$785$}\\
\hline
norm. emittance [mm$\times$mrad]  &   \multicolumn{3}{c|} {$0.08$} \\
\hline
slice energy spread $\sigma_E/E$ (\%)  &   \multicolumn{3}{c|} { $0.22$}\\ 
\hline
\multicolumn{4}{|c|}{\em Common FEL parameters}\\
\hline
undulator magnetic field [T] & \multicolumn{3}{c|}{$1$}\\
\hline
undulator period [cm] & \multicolumn{3}{c|}{$1.4$}\\
\hline
deflection parameter & \multicolumn{3}{c|}{$1.3$}\\
\hline
\multicolumn{4}{|c|}{\em Output FEL parameters}\\
\hline
FEL wavelength [nm]  & 2.0\\
\hline 
Twiss $\beta$ [m] &  6.16\\
\hline
Pierce parameter $\rho$ &  0.0018\\
\hline
inh. broad. gain length [m]  & 0.702\\
\hline
saturation power [MW]  & 861\\
\hline
saturation length [m]  & 17.7\\
\hline
coherence length [$\mu$m]  & 0.05\\
\hline
sat. power with slippage [MW]  & 826\\
\hline
\hline
\end{tabular}
\end{center}
\end{table}%

The efficiency of energy transfer from electrons to the electric field and so the gain of the process are summarized by the FEL Pierce parameter $\rho$,
\be
\label{scaling:3}
\rho=\frac{1}{4\pi\gamma}\sqrt[3]{2\pi\frac{J}{I_0}\left(\lambda_u K f_b(K)\right)^2}
\ee
where $f_b(K)=J_0(\xi)-J_1(\xi)$ is the planar undulator Bessel correction factor, of argument
$$
\xi=\frac{K^2}{4(1+K^2)}
$$
and $I_0=17$ kA the Alfven current.
The current is expressed in terms of the bunch root mean squared (rms) time duration $\sigma_\tau$ and of the bunch
charge $Q_b$ as
\be
\label{scaling:5}
I[A]=\frac{Q_b[C]}{\sigma_\tau[s]\sqrt{2\pi}}
\ee
The current density $J$ given by
\be
\label{scaling:6}
J\left[\frac{A}{m^2}\right]=
\frac{Q_b[C]}{\sigma_\tau[s]\sigma_x[m]\sigma_y[m](2\pi)^{3/2}}
\ee
where $\sigma_{x,y}[m]$ is the rms transverse size of the electron beam.

The gain length, determining the FEL growth rate, can be expressed in terms of $\rho$ as follows
\be
\label{scaling:7}
L_g = \frac{\lambda_u}{4\pi\sqrt{3}\rho}
\ee
The Pierce parameter gives an estimate of the natural bandwidth of the FEL,
$\Delta\omega/\omega\simeq\rho$ and rules also the power at saturation that writes
\be
\label{scaling:10}
P_S\simeq \sqrt{2}\rho P_E
\ee
$P_E$ being the electron beam power, linked to the peak current and
energy by the relation $P_E=E_{beam}~I$.
Then, the length of the undulator section needed to reach the saturated laser power -- the saturation length -- is
\be
\label{scaling:9}
L_S = 1.066 L_g\ln\left(\frac{9P_S}{P_0}\right)
\ee
where $P_0$ is the input seed power.

The effect of inhomogeneous broadening due to significant energy spread and emittance can be embedded in the previous formulae~\cite{Booklet, Dattoli}:
both contribute to increase the gain and saturation length. 
Furthermore, since the longitudinal beam size becomes comparable to the coherence length,
slippage corrections are taken into account resulting in an effective saturation power.
Table~\ref{tab:fel} shows the results obtained using simple and analytical scaling laws~\cite{Booklet,Dattoli}
to describe the FEL signal pulse evolution in terms of saturation length and saturation
power accounting for the beam emittance, the energy spread and the slippage corrections for the reported beam.

\section{Conclusions}

We employed the new ReMPII scheme to (numerically) generate a 1.3
GeV electron bunch with outstanding quality ($\sigma_E/E|{slice}=0.22\, \%$,
$\epsilon_n=80\, nm$ and compactness by using a single Ti:Sa laser
system and a preformed plasma channel of length 3.7 cm. To operate
with the ReMPI scheme a small portion of the Ti:Sa pulse has been
frequency doubled and tightly focused on the target to further ionize
the dopant and extract electron from the background. The main portion
passed through a time shaping device and after focusing by a large
F/\# paraboloid constituted the driving pulse(s) of the plasma wave.
The scheme takes advantage of the virtual absence of jitter between
the ionizing and driving pulses due to the usage of a single laser system.
This opens the possibility to precisely determine bot the bunch length
and energy of the final bunch. In the current setup numerical simulations
show that bunches with duration from $360\, as$ up to $2.2\, fs$ can be generated.

Analytical results of FEL performance, based onto a $2.2 \, fs$ long bunch, show that powerful $2\, nm$ $X$ radiation of peak power exceeding $0.8GW$  can be generated with state-of-the-art undulator parameters, provided that quality-preserving beam optics from plasma exit to the undulator is employed.

\section{Acknowledgments}
We thank Giuseppe Dattoli for his help and suggestions in estimating the FEL performance with
the electron beams discussed in the text. The research leading to these results has received funding from the
European Union's Horizon 2020 research and innovation program under
Grant Agreement No 653782 - EuPRAXIA. We also acknowledge financial
support from the ELI-ITALY Network funded by CNR.

\end{document}